\documentclass[12pt]{iopart}

\expandafter\let\csname equation*\endcsname\relax
\expandafter\let\csname endequation*\endcsname\relax

\usepackage[utf8]{inputenc}
\usepackage[english]{babel}
\usepackage[T1]{fontenc}
\usepackage{amsmath}
\usepackage{hyperref}
\usepackage[numbers]{natbib}
\usepackage{tikz}
\usepackage{graphicx}
\usepackage{dcolumn}
\usepackage{bm}
\usepackage{placeins}
\usepackage{amssymb}
\usepackage{amsthm}
\usepackage{mathtools}
\usepackage{tcolorbox}
\usepackage{soul}
\newtheorem{theorem}{Theorem}
\newtheorem{lemma}{Lemma}
\newtheorem{definition}{Definition}

\newtheorem{proposition*}{Proposition}
\newtheorem{observation}{Observation}

\begin{document}


\title{Robustness measures for quantifying nonlocality}

\author{Kyunghyun Baek$^{1}$, Junghee Ryu$^{*2,3}$, Jinhyoung Lee$^{\dagger 4}$}
\address{$^1$Institute for Convergence Research and Education in Advanced Technology, Yonsei University, Seoul 03722, Republic of Korea}%
\address{$^2$Center for Quantum Information R$\&$D, Korea Institute of Science and Technology Information (KISTI), Daejeon 34141, Republic of Korea}
\address{$^3$Division of Quantum Information, KISTI School, Korea University of Science and Technology, Daejeon 34141, Republic of Korea}%
\address{$^4$Department of Physics, Hanyang University, Seoul, 04763, Republic of Korea}%
\address{$^5$Center for Quantum Simulation, Korea Institute of Science and Technology (KIST), Seoul 02792, Republic of
Korea}%

\ead{$^*$junghee@kisti.re.kr, $^\dagger$hyoung@hanyang.ac.kr}


\begin{abstract}
We suggest generalized robustness for quantifying nonlocality { and derive its equivalence to the maximum violation ratio of Bell inequalities defined as vectors with non-negative elements.} We investigate its properties by comparing it with white-noise and standard robustness measures.
As a result, we show that white-noise robustness does not fulfill monotonicity under local { operations} and shared randomness, whereas the other measures do. 
To compare the standard and generalized robustness measures, we introduce the concept of inequivalence, which indicates a reversal in the order relationship depending on the choice of monotones. 
From an operational perspective, the inequivalence of monotones for resourceful objects implies the absence of free operations that connect them. Applying this concept, we find that standard and generalized robustness measures are inequivalent between even- and odd-dimensional cases up to eight dimensions. This is obtained using randomly performed CGLMP measurement settings in a maximally entangled state. 
This study contributes to the resource theory of nonlocality and sheds light on comparing monotones by using the concept of inequivalence valid for all resource theories.
\end{abstract}

\maketitle

\section{Introduction}

Bell's theorem, which is a seminal principle in quantum physics, states that certain quantum correlations cannot be simulated using local hidden variables (LHVs) model~\cite{Bell1964,Bell1966,Brunner2014}. This conflict arises when the predictions of quantum theory for the behavior of entangled systems contradict the assumptions of LHV.
Violations of Bell inequalities,  derived under the assumption of the LHVs, indicate that nature is incompatible with the LHV models, a phenomenon often referred to as (Bell) nonlocality. Many experiments have been conducted to confirm the nonclassicality of nature~\cite{Hensen2015,Giustina2015,Shalm2015}. 

Beyond its fundamental significance, the amount of Bell violations is considered an information-theoretic resource that enables various cryptographic tasks, such as device-independent (DI) quantum key distribution \cite{Acin2007,Pironio2009,Pirandola2020} and DI quantum random number generation \cite{Zhao2018,Pironio2010,Herrero2017}. Due to the impossibility of implementing these DI tasks using classical communications, nonlocality is considered as a valuable quantum resource that empowers quantum advantages unattainable by classical systems.

Quantum resources are primarily defined as the complement of a free set. 
{ In this framework, a quantum object is considered free if it can be realized using only the allowed set of physical operations. The collection of all such free objects forms the free set, while any object outside this set is regarded as a resource. An operation is classified as free if it cannot create a resource from any free object.} 
Resource theories provide a rigorous framework for comparing quantum resources and exploring available capabilities for information processing. Furthermore, by restricting free operations to currently feasible operations, researchers can analyze quantum advantages achievable in specific information tasks from an operational perspective. 
From these perspectives, many resource theories have been established for quantifying quantum resources, such as entanglement \cite{Vedral1998,Plenio2007,Horodecki2009}, coherence \cite{Johan2006,Baumgratz2014,Streltsov2017} and nonlocality \cite{de_Vicente2014,Brito2018,Camalet2020,Wolfe2020,Patryk2021}. 

Robustness measures are tools used to quantify quantum resources by accessing  the minimal noise that a quantum resource can withstand before becoming a free object, i.e. no quantum resource. When a quantum resource can tolerate only minimal noise before becoming a free object, it suggests that there are no significant advantages associated with using that resource in specific quantum tasks. A more recent development is the introduction of a general framework for quantifying quantum resources. This framework is applicable to any convex resource theory, in which a convex combination of free objects cannot generate a quantum resource \cite{Chitambar2019}. Robustness measures are versatile in this framework because they satisfy the necessary criteria for legitimate quantification and hold operational meanings in discrimination tasks \cite{Takagi2019_2,Oszmaniec2019,Takagi2019}.

In the resource theory of nonlocality, robustness measures are used to investigate the effect of noise on nonlocality. According to the type of noise, robustness measures are classified into white-noise \cite{Kaszlikowski2000,Acin2002} and standard \cite{Perez2008,Junge2010} robustness measures that quantify the tolerance to the addition of white and local noise, respectively. These robustness measures provide operational meaning, that is, how long the nonlocal correlations withstand realistic noise, such as detector efficiencies \cite{Eberhard1993}.

In this work, we aim to establish the family of robustness measures in the resource theory of nonlocality and explore their distinctions.
To this end, we suggest a generalized robustness that quantifies tolerance of the system to general noise { allowed under no-signaling conditions, i.e., any noise consistent with the axioms of special relativity} \cite{Barrett2005,Almeida2010}, { and derive its equivalence to the maximum violation ratio of Bell inequalities defined as vectors with non-negative elements}. 
We investigate various types of robustness measures, including white-noise, standard, and generalized robustness measures, to quantify nonlocality from a resource-theoretical point of view. 
Both the standard and the generalized robustness measures adhere to the monotonicity, which is a crucial requirement. This property ensures that a quantifier monotonically decreases under any free operation that cannot create a resource from a free object. Violation of monotonicity would result in the counter-intuitive behavior of a resource with the quantifier increasing under free operations.
However, we show that the white-noise robustness does not fulfill the monotonicity. We illustrate this with counter-examples, where the white-noise robustness increases under local operations and shared randomness (LOSR) \cite{de_Vicente2014}. This behavior occurs due to the dependence of the white-noise model on the number of dimension, as {LOSR} permits an increase in the number of outcomes during post-processing.

We also introduce the concept of `inequivalence' to systematically compare standard and generalized robustness measures. We define inequivalence as a behavior, where two monotones provide different order relations when used to compare two resourceful objects {\cite{Virmani2000,Liu2016,Zjawin2023}}.
In addition, we demonstrate that an inequivalent behavior implies the absence of free operations that connect objects. 
This concept provides a systematic method for numerically investigating the different behaviors of monotones with operational meaning. Applying this to our framework, we find that the standard and generalized robustness measures are inequivalent for cases, in which Alice and Bob share a maximally entangled state and randomly perform the optimal Collins–Gisin–Linden–Massar–Popescu (CGLMP) measurement settings \cite{Collins2002}. This is observed in all cases up to the eight-dimensional case. We extensively investigate the inequivalent behaviors of two- and three-dimensional cases for arbitrary states.

The remainder of this paper is organized as follows. In Sec. \ref{sec:2}, we review the resource theory of nonlocality and clarify the framework of our work. In Sec. \ref{sec:3}, we present the generalized robustness to quantify the nonlocality. In Sec. \ref{sec:4}, we show the non-monotonicity of white-noise robustness. In Sec. \ref{sec:5}, we introduce the concept of inequivalence and show that the standard and the generalized robustness measures are inequivalent by increasing dimensions of quantum systems. We summarize the findings of our study in Sec. \ref{sec:con}.

\section{Resource theory of nonlocality}\label{sec:2}

Consider the bipartite case, in which Alice and Bob share a quantum state $|\psi_{AB}\rangle$ on $\mathcal H_A\otimes \mathcal H_B$. In this scenario, they perform $m_A$- and $m_B$-outcome measurements $M_x^A$ and $M_y^B$ for $x,y\in\{1,2\}$, { where $m_A$ and $m_B$ are the number of measurement outcomes,} respectively. 
Importantly, these measurements are random, and neither Alice nor Bob have prior information about the quantum state. These results are presented in a simplified scenario but can be easily elaborated to cover multipartite cases with an arbitrary number of measurements. 
During each trial, Alice and Bob obtain measurement outcomes  $a_x \in\{1,2,...,m_A\}, b_y\in\{1,2,...,m_B\}$ given the choice of measurements $M_x^A$ and $M_y^B$, respectively. Consequently, Alice and Bob have access to the joint probability distribution $p(a_xb_y|xy)$ which satisfies the positivity $p(a_xb_y|xy)\geq 0$ and normalization conditions $\sum_{a_x,b_y}p(a_xb_y|xy)=1$ for all $x,y$.

In this scenario, each joint probability distribution is represented by the probability vector $\vec p_{xy} = (p(11|xy),...,p(m_A m_B|xy))^T$. Subsequently, we simply denote the all joint probability distributions as a direct-sum of probability vectors, $\vec P=\vec p_{11} \oplus \vec p_{12}\oplus \vec p_{21} \oplus \vec p_{22}\in [0,1]^{4m_Am_B}$. This representation satisfies the positivity and normalization conditions for each choice of measurements, as well as {\it no-signaling condition} 
\begin{align}\label{NS1}
\sum_{b_1} p(a_xb_1|x,1)-\sum_{b_2} p(a_xb_2|x,2)=0 \text{ for all $a_x,x$},\\
\sum_{a_1} p(a_1b_y|1,y)-\sum_{a_2} p(a_2b_y|2,y)=0 \text{ for all $b_y,y$}.
\end{align} 
This condition can be more compactly expressed as the linear equation:
\begin{align}\label{NS2}
G_{\mathcal{NS} }\vec P=0,
\end{align} 
where {  $G_{\mathcal{NS}}$} is $2(m_A+m_B) \times 4m_Am_B$ matrix corresponding to Eq. \eqref{NS1}.

The no-signaling condition implies that the choice of measurement made by party $A$ (or $B$) should not affect the marginal probabilities of the other party $p(b_y|y)$ (or $p(a_x|x)$). If this were allowed, the distant parties would be able to send signals based on their measurement choices. Such signaling is prohibited not only {in quantum mechanics but also in any physical theory, including relativity theory.}
Hence, we focus on the set of joint probability distributions that satisfy the no-signaling conditions as our general examination object in the resource theory of nonlocality, denoted by $\mathcal NS$.

\begin{figure}[t]
  \centering
    \includegraphics[width =7cm]{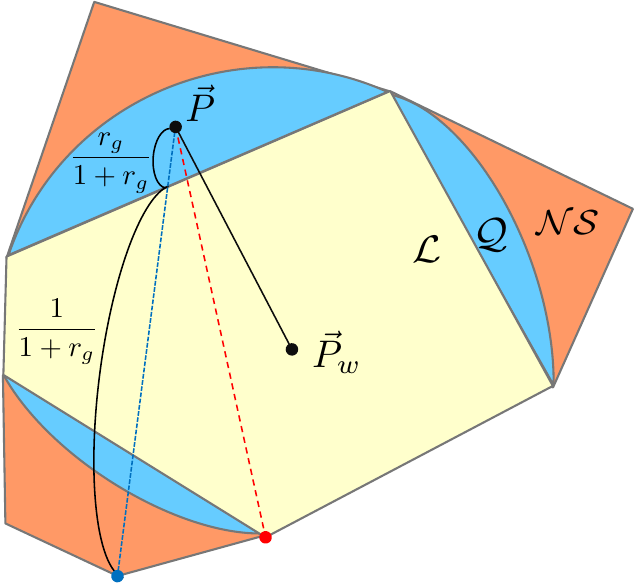}
    \caption{Geometrical aspects of robustness measures defined as the minimal noise that a nonlocal correlation can withstand before transitioning into a local correlation. Categories of robustness measures are based on added noise models and include white-noise (solid black line), local noise within $\mathcal{L}$ (dashed red line), and  general noise within $\mathcal{NS}$ (dotted blue line). The hierarchy $\vec P_{w}\in\mathcal{L}\subset\mathcal{Q}\subset\mathcal{NS}$ implies the relationship $R_{w}\geq R_S \geq R_G$. }\label{RT}
\end{figure}

A joint probability distribution $\vec P \in \mathcal NS$ is said to be local if it admits the {\it local hidden variable (LHV)} model; meaning, there exists a global probability distribution $l(a_1,a_2,b_1,b_2)$ that  generates the joint probability distribution \cite{Fine1982}
\begin{align}\label{Local1}
p(a_x,b_y|x,y)=\sum_{a_{\tilde x},b_{\tilde y}}l(a_1,a_2,b_1,b_2),
\end{align} 
where $\tilde x$ ($\tilde y$) is the counterpart of $x$($y$). The set of local probability distributions is denoted as $\mathcal L$, which is a local subset of $\mathcal NS$ as illustrated in Fig. \ref{RT}.
This condition of the LHV models can be written as follows:
\begin{align}\label{Local2}
G_{\mathcal L }\vec L=\vec P,
\end{align}
where $\vec L=(l(1,1,1,1),l(1,1,1,2),..., l(m_A,m_A,$ $m_B,m_B))^T\in [0,1]^{m_A^2 m_B^2}$ is a vector of a probability distribution, i.e. $\sum_{a_1,a_2,b_1,b_2} l(a_1,a_2,b_1,b_2)=1$, and $G_{\mathcal L }$ is the $4m_Am_B\times m_A^2m_B^2$ matrix corresponding to the local condition \eqref{Local1}.

A joint probability distribution that admits the LHV models does not provide {quantum advantages, particularly in DI scenarios, such as DI quantum random-number generation  \cite{Zhao2018,Pironio2010,Herrero2017} and DI quantum key distribution \cite{Acin2007,Pironio2009,Pirandola2020}}. Consequently, such distributions are considered free resources in the resource theory of nonlocality. In this framework, free operations are defined as processes that are incapable of converting local distributions into nonlocal ones. These operations were introduced in previous works within a strictly operational paradigm, known as {\it local operations and shared randomness (LOSR)} \cite{de_Vicente2014,Geller2014}. LOSR uses shared randomness and local probability transformations. More precisely, a joint probability distribution $\vec P$ can be freely transformed into 
$$\vec P' = q_0 \vec P_l + \sum_{k\geq 1} q_k O^L_k(\vec P),$$
where $q_k$ satisfies $\sum_{k\geq 0} q_k= 1$ while $\vec P_l \in \mathcal L$ and $O^L_k$'s are compositions of local operations, such as relabeling, input and output operations \cite{de_Vicente2014}.

\begin{table*}
  \centering
\begin{tabular}{ p{2.2cm}||p{9.3cm}p{2cm}p{1.8cm}  }
 \hline
  & Definition & Monotonicity & Convexity\\
 \hline
 \hline 
Generalized Robustness   & $ R_G(\vec P)= \min \{ r_g|\frac{\vec P+r_g \vec P_{ns}}{1+r_g}\in\mathcal L, \vec P_{ns}\in\mathcal{NS}\}$&\;\;\;\;\;\;\;Yes& \;\;\;\;\; Yes\\
Standard Robustness &  $R_S(\vec P)= \min \{ r_s|\frac{\vec P+r_s \vec P_{l}}{1+r_s}\in\mathcal L, \vec P_{l}\in\mathcal{L}\}$  &\;\;\;\;\;\;\;Yes  &\;\;\;\;\; Yes\\
White-noise Robustness & {\small $R_{w}(\vec P)= \min \{ r_{w}|\frac{\vec P+ r_w \vec P_{w}}{1+r_w}\in\mathcal L, \vec P_{w}=\frac{1}{m_Am_B}(1,...,1)^T\}$ }
&\;\;\;\;\;\;\;No& \;\;\;\;\;\;Yes\\
 \hline
\end{tabular}
  \caption{Definition of robustness measures and their properties.}
  \label{tab:1}
\end{table*}

\section{Robustness measures}\label{sec:3}

In this section, we will review white-noise and standard robustness measures and introduce generalized robustness in the nonlocality resource theory framework.
Each robustness represents the tolerance of a nonlocal correlation to the addition of specific noise models, such as white-noise, local noise, and no-signaling noise, respectively. Consequently, the operational relevance of each robustness depends on the characteristics of the noise present in a given experimental setting, {with no-signaling noise serving as an analytical tool rather than a commonly observed experimental phenomenon.} A more detailed discussion is provided in the following subsections.

\subsection{White-noise and standard robustness measures}

Robustness measures were adopted to investigate nonlocal correlations affected by specific noise models from an operational perspective. 
First, {\it white-noise robustness} was introduced to investigate the effects of adding the minimal amount of the maximally mixed state, needed for nonlocal correlations to vanish as illustrated in Fig. \ref{RT}. This can be formulated as follows { \cite{Kaszlikowski2000,Acin2002}}:
\begin{align*}
R_{w}(\vec P)=\min&\Big\{ r_w \geq 0 \Big|\frac{\vec P+r_{w} \vec P_{w}}{1+r_{w}}\in \mathcal L 
\text{ for } \vec P_{w}={ \frac{1}{m_Am_B}\left(1,...,1\right)^T} \Big\},
\end{align*}
where $m_A$ (or $m_B$) represents the number of outcomes for Alice's (or Bob's) measurement settings. $\vec P_{w}$ is a uniform vector in $4m_Am_B$-dimensional space, corresponding to a joint probability distribution, where possible outcomes are measured with equal probability because of the white-noise effect.


Another robustness is defined as the minimal addition of local noise that causes the nonlocal correlation to vanish, as shown in Fig. \ref{RT} \cite{Junge2010,Perez2008}.
This type of robustness is called {\it standard robustness} in the general resource theory framework \cite{Takagi2019,Chitambar2019}. Standard robustness refers to the minimal addition of free objects (local correlations) to a resourceful object (nonlocal correlation) before its resource (nonlocality) vanishes. It is formulated as { \cite{Perez2008,Junge2010}}:
\begin{align*}
R_S(\vec P)=\min\left\{ r_s\geq 0 \Big|\frac{\vec P+r_s\vec P_l}{1+r_s}\in \mathcal L \text{ for } \vec P_l\in \mathcal L \right\}.
\end{align*}
The minimal value $r_s$ corresponds to the maximal possible violation $\nu=2r_s+1$  by $\vec P$ of the Bell inequality, defined as \cite{Junge2010}
\begin{align}\label{eq:relevance_r_s}
\nu = \max_{\vec S} \frac{|\vec S \cdot \vec P|}{\max_{\vec P_l\in \mathcal L} |\vec S\cdot \vec P_{l}|},
\end{align}
where $\vec S$ is the Bell inequality acting on $\vec P$.
This relationship corresponds to that found in general convex resource theories, where standard robustness has an operational meaning; namely, the exact quantifier of the maximum advantage in discrimination tasks \cite{Takagi2019}.

\subsection{Generalized robustness}

In addition to standard robustness, {\it generalized robustness} is pivotal in various resource theories. It represents the minimal amount of general noise, permitted in the resource theory, that can be added to a resourceful object before its quantum resource vanishes. 
Applying this to the resource theory of nonlocality, we define the generalized robustness as
\begin{align*}
R_G(\vec P)=\min\left\{ r_g\geq 0 \Big|\frac{\vec P+r_g\vec P_{ns}}{1+r_g}\in \mathcal L \text{ for } \vec P_{ns}\in \mathcal{NS} \right\}.
\end{align*}

{

Analogous to the white-noise and standard robustness measures, the generalized robustness also admits a precise physical interpretation.
In the definition of $R_G$, $\vec P_{ns} \in \mathcal{NS}$ represents a more general type of noise allowed by physical theory than $\vec P_{l} \in \mathcal{L}$. Specifically, the no-signaling condition arises from the axioms of special relativity, which prohibit instantaneous signaling between parties (see Ref. \cite{Popescu2014} and references therein). In this sense, the generalized robustness provides a more conservative measure for quantifying tolerance to the addition of physically allowable noise.
}

{
As a generalized robustness also has an operational relevance in discrimination tasks within convex resource theories \cite{Takagi2019}, thus it is natural to question whether our generalized robustness is related to the violation ratio of Bell inequalities, similar to the standard robustness as defined in Eq. \eqref{eq:relevance_r_s}.
For this purpose, we reformulate the generalized robustness into the canonical form of linear programming in \ref{Appdx1:R_g}  and derive its dual characterization, with the proof provided in Appendix \ref{Appdx:proof_lamma_1}.
\begin{lemma}
    Let $\vec P$ is a direct-sum of probability vectors $\vec P = \vec p_{11}\oplus...\oplus \vec p_{22}\in [0,1]^{4 m_A m_B}$. Then the generalized robustness $R_G(\vec P)$ can be expressed as the following linear programming:
    \begin{equation}\label{eq:dual_rg}
        R_G(\vec P) +1 = \max \Big\{ \vec P^T \vec X | \vec P_l^T \vec X \leq 1 \text{ for all } \vec P_l \in \mathcal L, 
        (\vec X)_i \geq 0 \text{ for all }i.  \Big\},
    \end{equation}
    {  where $\vec X \in \mathbb{R}_+^{4{m_A m_B}}$.}
\end{lemma}
}

{
From the dual characterization and the definition of generalized robustness, we derive the following theorem:
\begin{theorem}
    The generalized robustness can be expressed as
    \begin{equation}\label{eq:rg_operational}
        R_G(\vec P) +1 = \max_{{\substack{\vec S \in \mathbb{R}_+^{4m_A m_B}}}} \frac{\vec S^T \vec P}{\max_{\vec P_l \in \mathcal L} \vec S^T \vec P_l}.
    \end{equation}
\end{theorem}
\begin{proof}
    We first prove that the left-hand side (LHS) of Eq. \eqref{eq:rg_operational} is greater than or equal to the right-hand side (RHS).  By the definition of generalized robustness, for any $\vec S \in \mathbb{R}+^{4m_A m_B}$, it holds that
    \begin{align*}
        \vec S^T \vec P  = \vec S^T \left( \left(1+R_G (\vec P)\right) \vec P_l^* - R_G (\vec P) \vec P_{ns}^*  \right)
        &\leq \left(1+R_G (\vec P)\right)  \vec S^T \vec P_l^*,\\
        &\leq \left(1+R_G (\vec P)\right) \max_{\vec P_l \in \mathcal{L}} \vec S^T \vec P_l,
    \end{align*}
    where $\vec P_l^*\in\mathcal{L}$ and $\vec P_{ns}^*\in \mathcal{NS}$ are the optimal vectors that provide the generalized robustness for the given $\vec P$. The first inequality follows from the non-negativity of $\vec S$.

    To prove LHS of Eq. \eqref{eq:rg_operational} is less than or equal to RHS of Eq. \eqref{eq:rg_operational}, we use the dual characterization of the generalized robustness, given in Eq. \eqref{eq:dual_rg}. Let $\vec X^*$ be the optimal vector that maximizes Eq. \eqref{eq:dual_rg}. Then, we have:
    \begin{align*}
        R_G(\vec P) + 1 &= \vec X^{*T}\vec P 
        \leq \frac{\vec X^{*T} \vec P}{\max_{\vec P_l \in \mathcal{L}} \vec X^{*T} \vec P_l } 
        \leq \max_{\vec S \in \mathbb{R}_+^{4 m_A m_B}} \frac{\vec S^{T} \vec P}{\max_{\vec P_l \in \mathcal{L}} \vec S^{T} \vec P_l },
    \end{align*}
    where the first inequality follows from the constraint $\vec X^T \vec P_l \leq 1$ for all $\vec P_l \in \mathcal{L}$ in the dual problem \eqref{eq:dual_rg}, and the second inequality follows from the fact that $\vec X^*$ is always included in $\mathbb{R}_+^{4m_A m_B}$.
\end{proof}

This result shows the generalized robustness corresponds to the maximum violation ratio achieved by $\vec P$ when Bell inequalities are restricted to vectors with non-negative elements, i.e., $\vec S\in \mathbb{R}_+^{4m_A m_B}$. 
Equivalently, this maximum violation ratio of these Bell inequalities by $\vec P$ indicates the tolerance to the addition of general type of noise allowed by the no-signaling condition. 

}

Standard and generalized robustness measures are applied selectively based on their properties in specific resource theories. For instance, standard robustness was initially proposed in the resource theory of entanglement \cite{Vidal1999}. Later, the theory was extended to encompass generalized robustness \cite{Harrow2003} for operational applications, such as one-shot entanglement dilution \cite{Brandao2010,Brandao2011}. 
In the resource theory of quantum coherence, the focus has been primarily on studying generalized robustness \cite{Piani2016, Napoli2016}. 
The preference for this measure arose because standard robustness exhibits divergence, namely, non-vanishing off-diagonal elements in the density matrix when mixed with incoherent states, represented by diagonal matrices \cite{Takagi2019}.

All the robustness measures discussed are summarized in Table \ref{tab:1}.
In the resource theory of nonlocality, these robustness measures can be numerically computed using linear programming via linear equations \eqref{NS2} and \eqref{Local2}
\footnote{In addition to robustness measures, the nonlocal content, also referred to as the nonlocal fraction or the 'EPR2' measure derived from the names of the authors who introduced it in \cite{Elitzur1992}, can be computed by using linear programming. The nonlocal content is defined as the minimum weight of the no-signaling distributions when a probability distribution is decomposed into local and no-signaling distributions, i.e., {  $\min\{0\leq \alpha\leq 1| \vec P = \alpha \vec P_{ns} + (1-\alpha)P_{l} \text{ for } \vec P_{ns}\in \mathcal{NS} \text{ and } \vec P_{l}\in \mathcal{L} \}$.} This measure has been proven to be monotone in the resource theory of nonlocality \cite{de_Vicente2014}.}.
For numerical investigations, we have reformulated all robustness measures into the canonical form of linear programming, as shown in \ref{Appdx1}.

\begin{figure}[t]
  \centering
    \includegraphics[width =7cm]{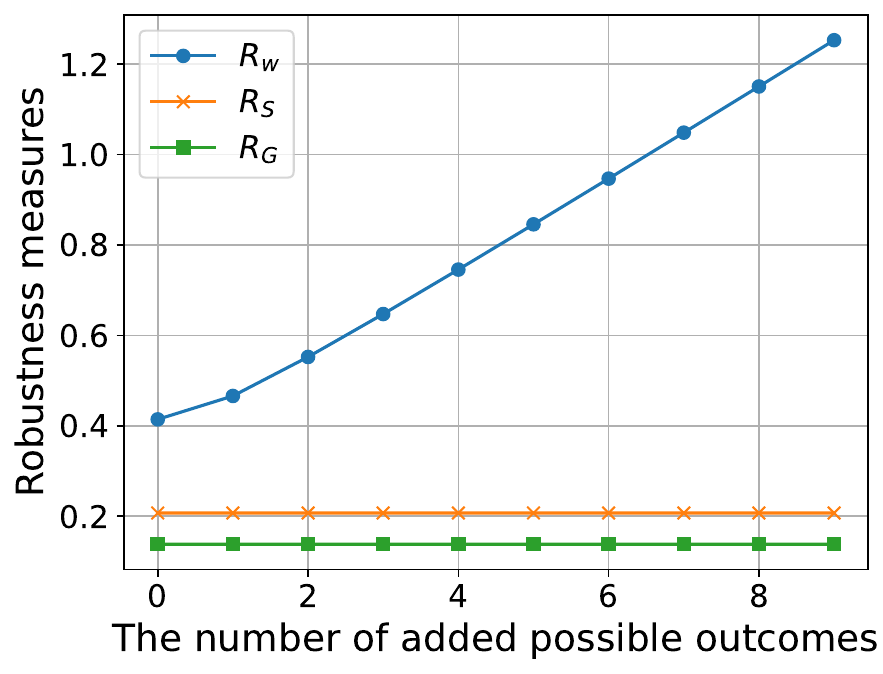}
    \caption{Non-monotonicity of white-noise robustness under LOSRs.}\label{NonMono}
\end{figure}

\section{Non-monotonicity of white-noise robustness}\label{sec:4}

In the resource theory framework, a quantifier should adhere to two key criteria: (i) faithfulness, meaning it should vanish if and only if the object is free, and (ii) monotonicity, meaning a quantifier should monotonically decrease under any free operations (LOSRs in our framework). 
By definition, all robustness measures are faithful. 
Furthermore, it has been shown that standard and generalized robustness measures fulfill monotonicity in all convex resource theories  \cite{Takagi2019}. Regardless, the monotonicity of white-noise robustness is not generally guaranteed \cite{Harrow2003}.

To verify the violation of the monotonicity of $R_{w}$, we provide counter-examples that demonstrate an increase in $R_{w}$ under the LOSRs. Consider output operations as a class of LOSRs that allows to artificially enlarge or merge possible measurement outcomes, as described in \cite{de_Vicente2014}. Output operations cannot generate nonlocal correlations from a correlation admitting LHV model. As a specific case, we assumed that $\mathcal E_k$ is an output operation that adds $k$ more outcomes to each measurement setting and assigns zero probability to the added outcomes, such that 
\begin{align*}
p_k(a_x,b_y|x,y) 
=  \begin{cases}
 p(a_x,b_y|x,y)       &  \text{for } a_x\leq m_A \text{ and } b_y\leq m_B\\
    0 &  \text{for } k \geq a_x> m_A \text{ or } k \geq b_y>m_B
  \end{cases}
\end{align*}
where $p_k(a_x,b_y|x,y)$ is the joint probability consisting of  {  $\mathcal E_k(\vec P)$}.

{ 
White noise is defined as the mixing of a uniform probability distribution with a given probability distribution, reflecting the fact that noise affects all possible outcomes uniformly. Consequently, the uniform probability distribution, $\vec P_{w}$, must be adjusted to account for an increased number of outcomes.  In other words, the effect of white noise on each probability becomes weaker as it is distributed across the newly introduced outcomes, thereby enhancing the robustness of nonlocal correlations against white noise.} For instance, if we apply the above operation $\mathcal E_k$ to $\vec P_{max}^{CHSH}$ giving the maximal violation of the Clauser-Horne-Shimony-Holt (CHSH) inequality for $m_A=m_B=2$, white-noise robustness increases strictly according to the number of added possible outcomes $k$. In contrast, the standard and generalized robustness measures are constant, as shown in Fig. \ref{NonMono}. This result confirms that the white-noise robustness is not monotonic under LOSRs.

{

On the other hand, $R_w$ can also increase under so-called output coarse graining operation, a process that merges measurement outcomes without reducing their total number \cite{de_Vicente2014}. In this case, the uniform distribution remains without adjustment as merging outcomes does not entail a reduction of output space. For example, consider a scenario that Alice and Bob share the maximally entangled state of two $d$-dimensional systems and perform CGLMP measurement settings randomly. The resulting joint probability distributions, denoted by $\vec{P}^{CGLMP}_d$, is given as \cite{Collins2002}
$$
p_{d}^{{CGLMP}}(a_x,b_y|x,y) = \frac{1}{2d^3 \sin^2[\pi (a_x-b_y+ \alpha_x + \beta_y)/d]}
$$
where the parameter are defined as $\alpha_1 = 0$, $\alpha_2 = 1/2$, $\beta_1=1/4$, and $\beta_2 = -1/4$.
For $d=2^n$, we apply an output coarse graining, denoted by $\mathcal K_n$, which maps $\vec{P}^{{CGLMP}}_{d=2^n}$ to $\vec{P}'$ by merging the outcomes of the joint probability distribution as follows:
\begin{align*}
 p'(a_x,b_y|x,y) 
=  \begin{cases}
\sum_{a'\in \mathcal{S}_{a_x}} \sum_{b'\in \mathcal{S}_{b_y}} p_{d=2^n}^{{CGLMP}}(a',b'|x,y) &
\text{for } a_x\leq 2^{n-1} \text{ and } b_y\leq 2^{n-1},\\
    0 & \text{for } a_x> 2^{n-1} \text{ or } b_y > 2^{n-1},
  \end{cases}
\end{align*}
where $p'(a_x,b_y|x,y)$ is the joint probability consisting of  $\mathcal K_n(\vec{P}^{CGLMP}_{d=2^n})$ for $\mathcal{S}_{a_x}=\{a_x, a_x+2^{n-1}\}$, $\mathcal{S}_{b_y}=\{b_y, b_y+2^{n-1}\}$.

In the joint probability distribution $\mathcal K_n(\vec{P}^{{CGLMP}}_{d=2^n})$, the nonzero probability corresponds to the one for $d=2^{n-1}$ accordingly
\begin{align}
p_{d=2^{n-1}}^{{CGLMP}}(a_x,b_x|x,y) 
= \sum_{\substack{a'\in \mathcal{S}_{a_x}\\b'\in \mathcal{S}_{b_y}}} p_{d=2^n}^{{CGLMP}}(a',b'|x,y).
\label{eq:merge}
\end{align}
The proof of Eq. \eqref{eq:merge} is provided in \ref{Appdx:proof}. This relation indicates $p_{d=2^n}^{{CGLMP}}$ is reduced to $p_{d=2^{n-1}}^{{CGLMP}}$ with zero probabilities assigned to the redundant outcomes. 
We numerically demonstrate 
$$R_w(\vec P^{CGLMP}_{d=4})=0.448< R_w(\mathcal{K}_2(\vec P^{CGLMP}_{d=4}))= 0.552$$
implying the non-monotonicity of white-noise robustness without artificially increasing the number of outputs. Hence, the white-noise robustness is not a monotone even if the uniform distribution is preserved. 
Table \ref{tab:2} also shows the non-monotonicity of $R_w$ under output coarse graining by applying $\mathcal K_n$ sequentially to $\vec P^{CGLMP}_{d=16}$, while $R_S$ and $R_G$ show monotonic behaviors.

{  Consequently, more caution is needed when using the white-noise robustness to compare the amount of nonlocality.}
However, it is important to note that the non-monotonicity of white-noise robustness does not undermine its ability to certify {  the existence of nonlocality.  Namely, the whit-noise robustness still satisfies the faithfulness.} Despite this non-monotonic behavior, it remains a reliable indicator for detecting nonlocality and retains its operational meaning as the tolerance to the addition of white noise.

\begin{table}
  \centering
\begin{tabular}{c||c c c }
 \hline
  & $R_w$ & $R_S$ & $R_G$\\
 \hline
 \hline 
$\vec P^{CGLMP}_{d=16}$   &0.4755 & 0.2229 &0.1585 \\
$\mathcal{K}_4(\vec P^{CGLMP}_{d=16})$&  0.6216  & 0.2071  & 0.1554 \\
$\mathcal{K}_3\circ\mathcal{K}_4(\vec P^{CGLMP}_{d=16})$&  1.0243  & 0.2071  & 0.1494 \\
$\mathcal{K}_2\circ\mathcal{K}_3\circ\mathcal{K}_4(\vec P^{CGLMP}_{d=16})$& 1.7673   & 0.2071  & 0.1381 \\
 \hline
\end{tabular}
  \caption{ Non-monotonicity of white-noise robustness under output coarse graining.}
  \label{tab:2}
\end{table}

}



\section{Inequivalence of robustness measures}\label{sec:5}

In this section, we will define the inequivalence of monotones and conduct extensive numerical investigations of inequivalence between the standard and generalized robustness measures. 
\footnote{In the CHSH scenario, it has been shown in Appendix E of Ref. \cite{Goh2018} that various visibilities against white noise, local noise, and no-signaling noise are monotonically increasing functions of the CHSH violation. Therefore, in this subsection, we will investigate the inequivalence between the standard and the generalized robustness measures, focusing more on high-dimensional cases.}

\subsection{Inequivalence of monotones}

One of the primary purpose of quantifying resources is to investigate and compare their quantities. However, even when quantifiers satisfy both faithfulness and monotonicity criteria, they can yield different behaviors when used to compare the resourcefulness of objects. 
\begin{definition}[Inequivalence]
Quantifiers $Q_1$ and $Q_2$ are considered {\it inequivalent} for objects $o_1$ and $o_2$,  if quantifiers provide different order relationships for objects $o_1$ and $o_2$ such that $Q_1(o_1)>Q_1(o_2)$ and $Q_2(o_1)< Q_2(o_2)$.
\end{definition}
\noindent
{In this context, an object refers to a conceptually defined entity with physical or operational relevance within the framework of a resource theory. In our case, experimentally obtained probability distributions serve as the objects, and their resourcefulness corresponds to the amount of nonlocality they exhibit. }

\begin{figure}[t]
  \centering
    \includegraphics[width =7cm]{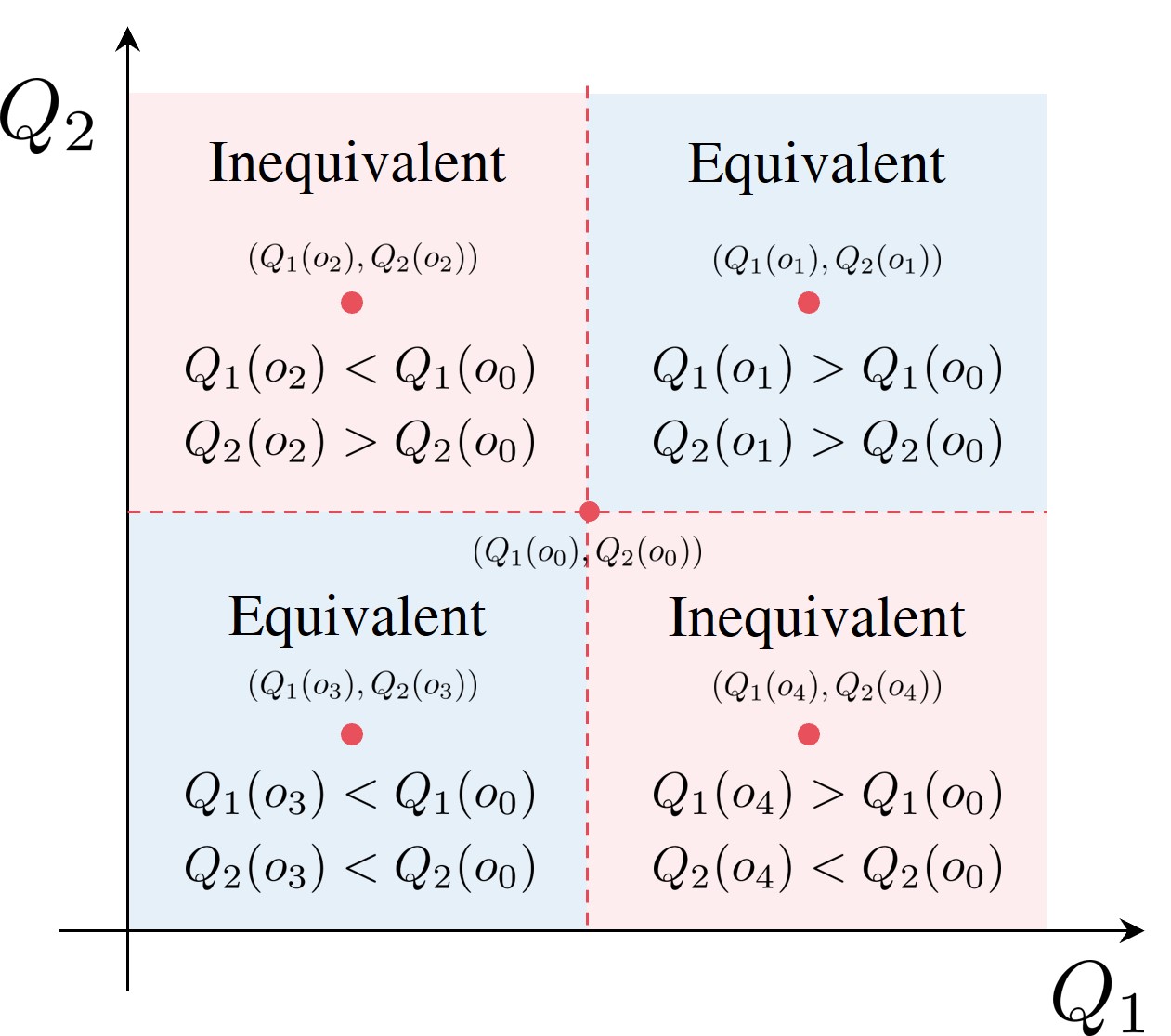}
    \caption{ Plot of a monotone $Q_2$ against another monotone $Q_1$ for a reference object $o_0$ compared with other objects. Points $(Q_1(o_i),Q_2(o_i))$ located in the first and the third quadrants (blue) indicate equivalence between $Q_1$ and $Q_2$, while those in the second and the fourth quadrants (red) indicate inequivalence.}\label{Equiv}
\end{figure}

Quantifiers satisfying the faithfulness and monotonicity criteria are called {\it monotones}; they decrease monotonically under free operations.\footnote{{ We use the term 'monotones' to emphasize that Observation \ref{obs} applies to any quantifier satisfying monotonicity. In contrast, we refer to robustness as 'measures' to follow conventions and to distinguish that they are not only monotones but also satisfy convexity conditions and are associated with specific operational interpretations. }}
{
The inequivalence between monotones has been extensively investigated in various resource theories, including entanglement \cite{Virmani2000}, coherence \cite{Liu2016}, and steering \cite{Zjawin2023}. 
Beyond its mathematical significance, this inequivalence holds important operational implications as follows:
}
{
\begin{observation}\label{obs}
    If some monotones $Q_1$ and $Q_2$ are inequivalent for objects $o_1$ and $o_2$, no free operation can transform one object into another.
\end{observation}
}
\begin{proof}
If we can generate an object $o_1$ from another object $o_2$ via a free operation $\mathcal F$, $o_1=\mathcal F(o_2)$, then any monotones $Q$ should establish equivalent relations such that $Q(o_1)\geq Q(o_2)$, in accordance to the monotonicity criterion. Thus, by the contrapositive of this statement, inequivalent behaviors for objects $o_1$ and $o_2$ imply no free operation connecting them.
\end{proof}

In general, determining whether it is possible to transform one object into another by a free operation is not straightforward. However, according to this observation, the inequivalence of two objects provides confirmation of the absence of free operations connecting them.
In our framework, if monotones such as the standard and generalized robustness measures are inequivalent for the vectors $\vec P_1$ and $\vec P_2$, then there is no LOSR $\mathcal E$ satisfying  $\mathcal E(\vec P_1)=\vec P_2$ or  $\mathcal E(\vec P_2)=\vec P_1$.  We note, however, the equivalence of monotones for $\vec p_1$ and $\vec p_2$ does not imply the existence of a LOSR connecting them.

To examine the inequivalence of monotones $Q_1$ and $Q_2$ for many objects simultaneously, we have plotted $Q_2$ against $Q_1$ in Fig. \ref{Equiv}. Subsequently, based on the object of interest $o_0$, objects located in the upper right and lower left corners are considered equivalent, whereas objects located in the upper left and lower right corners are considered inequivalent. We apply this method to investigate the inequivalence of the standard and generalized robustness measures in the following section.

\subsection{Inequivalence of robustness measures}

We also explore the difference between the robustness measures $R_S$ and $R_G$ by investigating their inequivalence in specific examples. 
{  In the following examples, we denote measurement outcomes as $a_x, b_y \in \{0, \dots, d-1\}$ for consistency with the CGLMP measurement settings in \cite{Collins2002}.}
We assume that Alice and Bob share a maximally entangled state: 
$$|\Psi_d\rangle_{AB}=\frac{1}{\sqrt d} \sum_{i=0}^{d-1} |ii\rangle_{AB}.$$
Alice and Bob then randomly perform measurements $A_x$ and $B_y$ for $x,y=1,2$ with nondegenerate eigenstates in $d$-dimensional Hilbert space, respectively.
\begin{equation}\label{eq:measure}
\begin{aligned}
&|a_x \rangle_{A,x}=
 \begin{cases}
\frac{1}{\sqrt{d_{\text{eff}}}} \sum_{j=0}^{d_{\text{eff}}-1} \exp\left(i \frac{2\pi}{d_{\text{eff}}} j(a_x+\alpha_x)\right) |j\rangle_A &\text{ for }k<d_{\text{eff}} \\
    |a_x\rangle_A   &\text{ for } d_{\text{eff}} \leq a_x < {  d}
  \end{cases} \\
&|b_y\rangle_{B,y}=
\begin{cases}
\frac{1}{\sqrt{d_{\text{eff}}}} \sum_{j=0}^{d_{\text{eff}}-1} \exp\left(i \frac{2\pi}{d_{\text{eff}}} j(-b_y+\beta_y)\right) |j\rangle_B & \text{for }l<d_{\text{eff}} \\
    |b_y\rangle_B &  \text{for } d_{\text{eff}} \leq  l < {  d}
  \end{cases}
\end{aligned}
\end{equation}
with effective dimensions $d_{\text{eff}}$, where $\alpha_1=0,\alpha_2=1/2$, $\beta_1=1/4$, and $\beta_2=-1/4$.

\begin{figure}
  \centering
    \includegraphics[width =7cm]{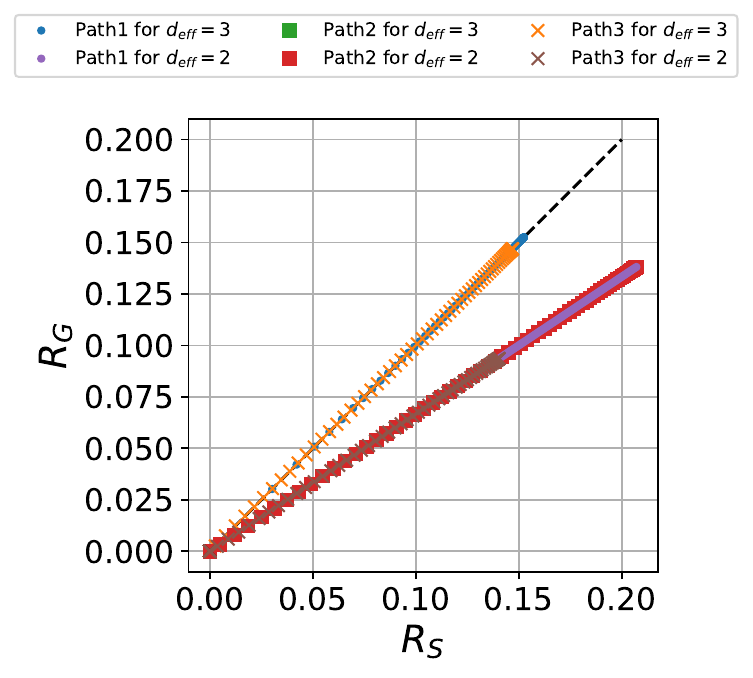}
    \caption{Inequivalence between standard ($R_S$) and generalized ($R_G$) robustness measures. All values along the paths 1, 2 and 3 for each $d_\text{eff}=2$ and $3$ are located on the lines with slopes of 1.5 and 1, respectively.}\label{path_3d}
\end{figure}

The measurement settings for $d_\text{eff}={  d}$ were used in \cite{Collins2002,Kaszlikowski2000} as the optimal measurements for achieving the maximal Bell violation, obtained through numerical investigations. We introduce the effective dimension to consider more general cases, where the measurements are performed in the subspace of the entangled states. 
In these settings, $d$ represents the degree of entanglement of  $|\Psi_d\rangle_{AB}$, while $d_\text{eff}$ represents the incompatibility of the measurement settings. This approach allows us to explore scenarios in which highly entangled states are measured using less incompatible measurements, i.e. $d>d_{\text{eff}}$. (See examples in Fig. \ref{path_3d}.)

\begin{figure}
  \centering
    \includegraphics[width =7cm]{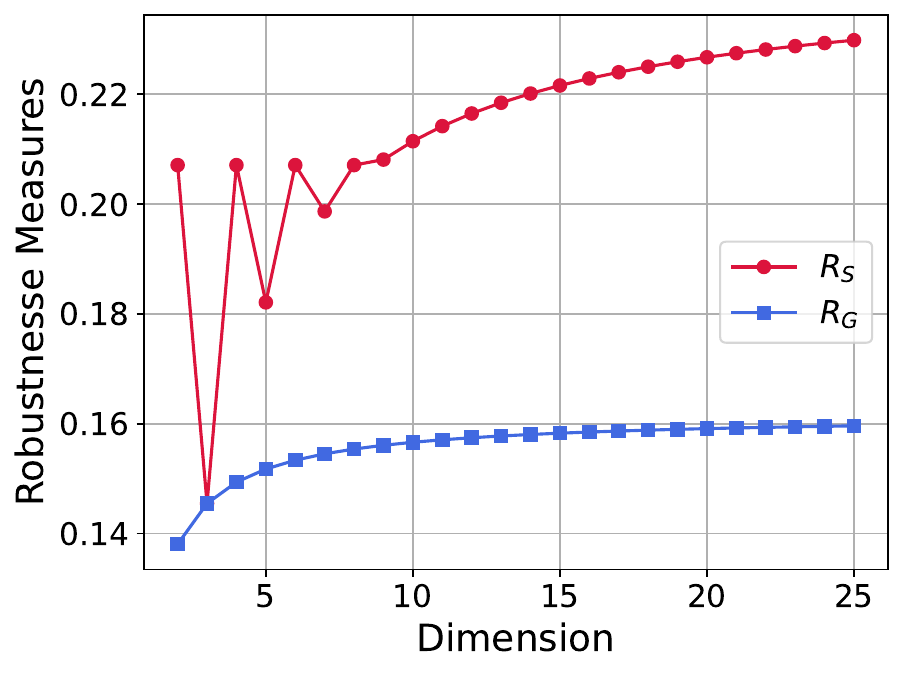}
    \caption{Standard ($R_S$) and generalized ($R_G$) robustness measures for sharing maximally entangled states and performing the optimal CGLMP measurement settings for dimensions ranging from $d=2$ to $25$.  Inequivalence between standard and generalized robustness measures is observed between even and odd dimensions up to $d=8$.}\label{Ineq_d_1}
\end{figure}

\begin{figure*}[t]
  \centering
    \includegraphics[width =17cm]{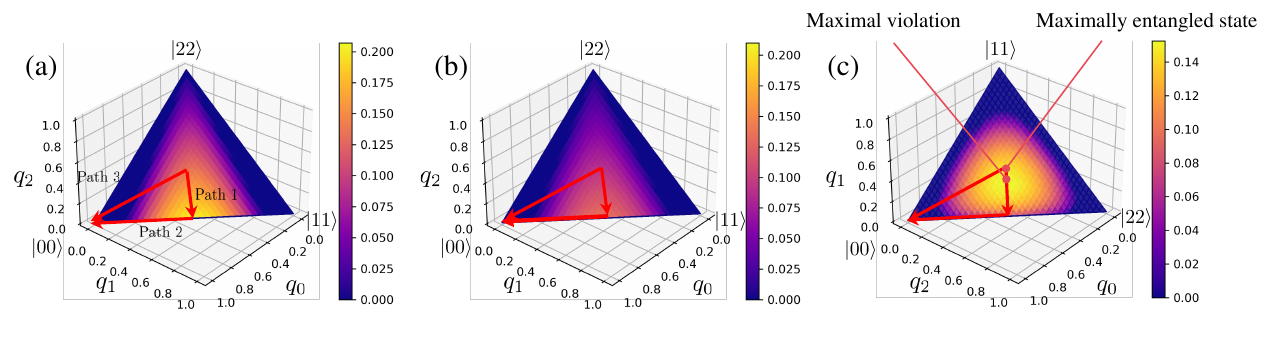}
    \caption{Standard ($R_S$) and generalized ($R_G$) robustness measures for $|\psi\rangle=\sqrt{q_0}|00\rangle+\sqrt{q_1}|11\rangle+\sqrt{q_2}|22\rangle$ with measurement settings for $d_\text{eff} = 2,3$. (a) Standard and generalized robustness measures for $d=d_\text{eff}=3$ have the same values for all parameters. (b) Standard and (c) generalized robustness measures for $d=3$ and $d_\text{eff}=2$ have different values, but they show a consistent ratio.}\label{R_d3}
\end{figure*}

For $d=d_{\text{eff}}$, higher dimensions, which imply higher entanglement and incompatibility, are expected to result in larger nonlocality. 
{ 
Namely, nonlocality is intuitively expected to increase with dimension. The generalized robustness aligns with this expectation, exhibiting a monotonic increase for 
$d$ from 2 to 25.
} However, the standard robustness yields inconsistent results: it remains constant at 0.207107 for even dimensions up to ${  d}=8$ and increases for odd dimensions from 0.145489 to 0.198678 up to ${  d}=7$ as shown in Fig. \ref{Ineq_d_1}. Thus, they exhibit inequivalence with an increase in the number of dimensions up to eight. Consequently, we can confirm that no LOSR operation transforms the probability distributions from odd- to even-dimensional cases and vice versa. This result contrasts with the case where an entangled state can be transformed from $|\Psi_d\rangle$ to $|\Psi_{d'}\rangle$ for $d>d'$ via local operations and classical communications \cite{Nielsen1999}.

The inequivalence stands out most distinctively between the two- and three-dimensional cases. Thus, we further examine the inequivalence of these cases more extensively by considering $|\psi\rangle = \sqrt{q_0}|00\rangle+\sqrt{q_1}|11\rangle+\sqrt{q_2}|22\rangle$ for positive values $q_0+q_1+q_2=1$ and the measurement settings for $d_\text{eff}=2,3$ as defined in Eq. \eqref{eq:measure}.
For $d=2$, the generalized and standard robustness measures exhibit similar behaviors, only with different scales, as shown in Fig. \ref{R_d3} (a) and (b). They both had maximal values for $|\Psi_2\rangle$. For $d=3$, the standard and the generalized robustness measures had the same values, as shown in Fig. \ref{R_d3} (c). In this setting, they attain their maximum value for the partially entangled state $(|00\rangle+\gamma|11\rangle+|22\rangle)/\sqrt{n}$, where $\gamma\simeq 0.7923$ and $n=2+\gamma^2$. This maximal violation condition is consistent with the result in \cite{Collins2002}.

We further plot the generalized robustness against the standard robustness for $d_\text{eff}=2,3$ in Fig. \ref{path_3d}, by changing the states along the paths 1, 2 and 3 as described in Fig. \ref{R_d3}. For $d_\text{eff}=2$, we change the states consistently along the paths 1, 2, and 3 from $|\Psi_3\rangle$ to $|\Psi_2\rangle$, $|\Psi_2\rangle$ to $|00\rangle$ and $|\Psi_3\rangle$ to $|00\rangle$, respectively. Similarly, we vary the states for $d_\text{eff}=3$ from $|\Psi_3\rangle$ to $(|00\rangle+|22\rangle)/\sqrt{2}$, $(|00\rangle+|22\rangle)/\sqrt{2}$ to $|00\rangle$ and $|\Psi_3\rangle$ to $|00\rangle$, respectively (See  \ref{Appdx2} for more details). 
We include the maximum values along these paths. Consequently, for each case of $d_\text{eff}=2,3$, all the points fall along the lines with slopes of 1.5 and 1. This shows that $R_G$ and $R_S$ are equivalent in each measurement setting. However, by comparing the probability vectors obtained across the measurement settings for $d_\text{eff}=2, 3$, we identify the inequivalence between $R_G$ and $R_S$. This result suggests that the inequivalence depends more on the measurement settings than on the type of state in this scenario.

\section{Conclusion}\label{sec:con}

We propose a generalized robustness for quantifying nonlocality in the resource-theoretic framework { and derive its equivalence to the maximum violation ratio of Bell inequalities defined as vectors with non-negative elements}. We then investigate the overall properties of white-noise, standard and generalized robustness measures. Our results show that the white-noise robustness is not monotonic, as it increases under LOSRs in our counter-example. In contrast, standard and generalized robustness measures are monotonic as established in the convex resource theory \cite{Chitambar2019}. 

Additionally, we introduce the concept of inequivalence, which signifies different order relationships when comparing the amount of resource for quantum objects. In our framework, this corresponds to the amount of nonlocality for probability vectors. The concept of inequivalence holds operational significance, implying that two quantum objects cannot be transformed from one to another if they exhibit inequivalence for specific monotones. When we apply this concept to compare the standard and generalized robustness measures, we observe inequivalence for probability vectors obtained by sharing a ${  d}$-dimensional maximally entangled state and randomly performing optimal measurement settings. This inequivalence persists between odd and even dimensions up to ${  d}=8$. Our numerical examination of two- and three-dimensional cases reveal that the degree of inequivalence is more dependent on the measurement settings. 

Our method of comparing monotones using the concept of inequivalence has potential applications in various resource theories, including coherence, entanglement, and nonclassicality.
Furthermore, it can be applied for verifying the impossibility of connecting quantum objects, an important issue in the classification of quantum resources like W and GHZ states within entanglement theory.
However, verifying this impossibility is not straightforward, as it requires consideration of all the possible free operations. 
Our method yields a conclusive result, indicating the absence of free operations only when we observe the inequivalence among  quantum objects. This method can be utilized to explore different monotone behaviors and the impossibility of connecting quantum objects via free operations.

\ack
{ We thank Seungjin Lee for helpful discussions. This work was supported by the National Research Foundation of Korea (NRF) grant funded by the Korea Government (MSIT) (Grant No. 2021M3E4A1038213, Grant No. 2022M3E4A1077094, ‌Grant No. 2022M3H3A106307411, ‌Grant No. RS-2023-NR119924, Grant No. RS-2023-NR119931, Grant No. RS-2023-00281456). JR was supported by Korea Institute of Science and Technology Information (KISTI) (No. K25L1M3C3). KB was supported by the Ministry of Science, ICT and Future Planning (MSIP) and the Institute of Information and Communications Technology Planning and Evaluation grant funded by the Korean Government (2019-0-00003, ‘Research and Development of Core Technologies for Programming, Running, Implementing and Validating of Fault-Tolerant Quantum Computing System’). 

\appendix

\section{Robustness measures in canonical form of linear programming}\label{Appdx1}

Numerical optimizations are essential for computing robustness measures, which are defined as the minimum amount of noise, required to eliminate quantum properties, —specifically, nonlocal correlations in this study. 

For convex quantum resources, robustness measures can be computed using convex optimizations, including linear and semidefinite programming. 
In this study, we chose linear programming because of its compatibility with formulating constraints as linear equalities or inequalities. 
This approach leverages the convex polytope nature of the feasible region. 
Therefore, we reformulated the robustness measures into the canonical form of linear programming to simplify computation.

\subsection{Generalized robustness}\label{Appdx1:R_g}

In this subsection, we rewrite generalized robustness as the canonical form of linear programming. {  For the simplicity, we assume $m_A=m_B=d$ throughout Appendix A, but the result can be straightforwardly extended to arbitrary $m_A$ and $m_B$.} Generalized robustness can be expressed as
\begin{align*}
R_G(\vec P)&=\min\left\{ r_g\geq 0 \Big|\frac{\vec P+r_g\vec P_{ns}}{1+r_g}\in \mathcal L \text{ for } \vec P_{ns}\in \mathcal{NS} \right\}\\
&=\min\Big\{ r_g\geq 0 \Big|{\vec P+r_g\vec P_{ns}}\in ({1+r_g}) \mathcal L 
\text{ for } \vec P_{ns}\in \mathcal{NS} \Big\}\\
&=\min\Big\{ r_g\geq 0 \Big|{r_g\vec P_{ns}} = ({1+r_g})\vec P_l - \vec P 
\text{ for } \vec P_l \in \mathcal L \text{ and } \vec P_{ns}\in \mathcal{NS} \Big\}\\
&=\min\left\{ r_g\geq 0 \Big| ({1+r_g})\vec P_l - \vec P \in \mathcal C_{\mathcal{NS}} \text{ for } \vec P_l \in \mathcal L \right\}
\end{align*}
where $\mathcal C_{\mathcal{NS}}$ is the cone generated by $\mathcal{NS}$, $\mathcal C_{\mathcal{NS}} :=\{r\vec P \;| r\in \mathbb{R}_+, \vec{P} \in \mathcal{NS}\}$. By replacing $\vec P_l' \equiv (1+r_g)\vec P_l$, we reduced it to 
\begin{align*}
R_G(\vec P) + 1 = \frac{1}{4} \min\left\{ \vec u^T \vec P'_l \Big| \vec P'_l - \vec P \in \mathcal C_{\mathcal{NS}} \text{ for } \vec P'_l \in \mathcal{C}_{\mathcal L} \right\},
\end{align*}
where $\vec u = (1,1,...,1)^T\in \mathbb{R}^{4d^2}$ is a uniform vector, 1/4 is multiplied for normalization, and $\mathcal C_\mathcal{L}$ is the cone generated by $\mathcal L$. 
Here, $\vec P_l'$ is in the cone of the local polytope implying that it should satisfy the no-signaling condition. Similarly, $\vec P$ is a joint distribution derived from Born's rule, also implying that it should adhere to the no-signaling condition. Thus, the non-negativity, i.e., $\vec P'_l - \vec P\geq 0$, is sufficient to guarantee $\vec P'_l - \vec P \in \mathcal C_{\mathcal{NS}}$. However, $\vec P'_l \in \mathcal C_{\mathcal{L}}$ holds true if and only if it can be written as $\vec P_l'=G_{\mathcal{L}}\vec L$ as shown in Eq. \eqref{Local2}, where each element of $\vec L$ is non-negative. 
Hence, by using the above linear conditions, the generalized robustness can be rewritten as 
\begin{align*}
R_G(\vec P)+1 = \frac{1}{4} \min \Big\{ \vec u^T (G_{\mathcal L} \vec L)| (G_{\mathcal L} \vec L-\vec P)_i\geq 0,  l(a_1,a_2,b_1,b_2)\geq 0 \text{ for } \vec L\in R_{+}^{d^4} \Big\},
\end{align*}
which is the canonical form of linear programming
\begin{align*}
\min \;\;\;\;&  \frac{1}{4}\vec u^TG_\mathcal{L}\vec L-1\\
s.t. \;\;\;\;& (G_{\mathcal L} \vec L-\vec P)_i \geq 0 \text{ for all } i \\
&l(a_1,a_2,b_1,b_2)\geq 0 \text{ for all } a_1,a_2,b_1,b_2.
\end{align*}


\subsection{Standard robustness}

Analogous to the treatment of generalized robustness, the standard robustness can be expressed as follows:
\begin{align*}
R_S(\vec P) + 1 = \frac{1}{4} \min\left\{ \vec u^T \vec P'_l \Big| \vec P'_l - \vec P \in \mathcal C_{\mathcal{L}} \text{ for } \vec P'_l \in \mathcal{C}_{\mathcal L} \right\}.
\end{align*}
By employing linear conditions for the local polytopes, this expression can be further reformulated as
\begin{align*}
R_S(\vec P) + 1 = & \frac{1}{4} \min \Big\{ \vec u^T G_{\mathcal L} \vec L \Big| G_\mathcal{L} \vec L  - \vec P = G_\mathcal{L} \vec L' \\ 
&\text{ for } l(a_1,a_2,b_1,b_2), l'(a_1,a_2,b_1,b_2) \geq 0 \Big\}.
\end{align*}
Formulating this in the context of linear programming yields
\begin{align*}
\min \;\;\;\;&  \frac{1}{4} \vec u^T G_\mathcal{L} \vec L-1\\
s.t. \;\;\;\;& G_{\mathcal L} (\vec L - \vec L') = p , \\
&l(a_1,a_2,b_1,b_2)\geq 0 \text{ and } l'(a_1,a_2,b_1,b_2)\geq 0 \\
&\text{for all } a_1,a_2,b_1,b_2.
\end{align*}

\subsection{White-noise robustness}

Similarly, white-noise robustness can be reformulated as follows:
\begin{align*}
R_w(\vec P) = \frac{1}{4} \min\Big\{r_w & \Big| (1+r_w) \vec P_l - \vec P =  r_w \vec P_w, \text{ for } \vec P_l \in {\mathcal L}  \text{ and }\vec P_w = \frac{1}{d^2} (1,...,1)^T \in \mathbb{R}_+^{4d^2} \Big\},
\end{align*}
which can be further expressed as
\begin{align*}
R_w(\vec P) +1 = \frac{1}{4} \min\Big\{\vec u^T \vec P_l' & \Big| (\vec P_l' - \vec P)_i = (\vec P_l' - \vec P)_j\geq 0 \text{ for all } i,j \text{ and } \vec P_l' \in \mathcal{C}_{\mathcal L} \Big\}.
\end{align*}
Finally, this formulation leads to the following canonical form of linear programming:
\begin{align*}
\min \;\;\;\;&  \frac{1}{4} \vec{u}^T G_\mathcal{L} \vec L -1\\
s.t. \;\;\;\;& (G_{\mathcal L} \vec L - \vec P)_i= (G_{\mathcal L} \vec L- \vec P)_j \geq 0  \text{ for all } i,j.
\end{align*}

{
\section{Proof of Lamma 1}\label{Appdx:proof_lamma_1}

Dual of a linear programming is well-defined, and strong duality holds if the primal problem is feasible according to the Slater's condition. \cite{Boyd2004}. 
Consider a linear programming problem in the following form: 
\begin{align*}
    \min \;\;\;\;&  \vec b^T \vec y\\
    s.t. \;\;\;\;& A^T \vec y  \geq \vec c \text{ and }  \vec y \geq 0 .               
\end{align*}
The dual of this linear programming is given as \cite{Matouek2006}
\begin{align*}
    \max \;\;\;\;&  \vec c^T \vec x\\
    s.t. \;\;\;\;& A \vec x  \geq \vec b \text{ and }  \vec x \geq 0 .
\end{align*}

For the case of generalized robustness, the dual of the corresponding linear programming problem can be expressed as
\begin{align*}
\max \;\;\;\;&  \vec P^T \vec X - 1\\
s.t. \;\;\;\;& \left[ G_{\mathcal L}^T (\frac{1}{4}\vec u - \vec X)\right]_i \geq 0 \text{ for all } i \\
&(\vec X)_i \geq 0 \text{ for all } i.
\end{align*}
Here, the first inequality is equivalent to the following constraint $\vec L^T G_{\mathcal L}^T (\frac{1}{4} \vec u - \vec X) \geq 0 $ for all probability vectors $\vec L$, as $\vec L$ can be chosen as a basis for each entry. 
By using $\vec P_l = G_{\mathcal L} \vec L$ and $\vec P_l^T \vec u = 4$, this constraint is reduced to $\vec P_l^T \vec X \leq 1$ for all $\vec P_l \in \mathcal L$. 
Additionally, the strong duality holds due to the existence of a feasible solution, such as $\vec X = 1/2 \vec u$. As a result, this strong duality implies that the generalized robustness can be reformulated into its dual characterization as defined in Eq. \ref{eq:dual_rg}.

}

{
\section{Proof of Eq. \eqref{eq:merge}}\label{Appdx:proof}
To prove Eq. \eqref{eq:merge}, we expand the right hand side of Eq. \eqref{eq:merge} by the definition of $p_{d=2^n}^{\text{CGLMP}}(a',b'|x,y)$ to
\begin{align*}
& \sum_{a'\in \mathcal{S}_{a_x}} \sum_{b'\in \mathcal{S}_{b_y}}  p_{d=2^n}^{\text{CGLMP}}(a',b'|x,y) \\
&= p_{d=2^n}^{\text{CGLMP}}(a_x,b_y|x,y) + p_{d=2^n}^{\text{CGLMP}}(a_x+2^{n-1},b_y|x,y) \\
&\;+ p_{d=2^n}^{\text{CGLMP}}(a_x,b_y+2^{n-1}|x,y) + p_{d=2^n}^{\text{CGLMP}}(a_x+2^{n-1},b_y+2^{n-1}|x,y)\\
&=  \frac{2}{2^{3n+1} \sin^2[\theta_{a_x,b_y}/2^{n}]}  + \frac{1}{2^{3n+1}\sin^2[\pi/2+ \theta_{a_x,b_y}/2^{n}]}\\
&\; + \frac{1}{2^{3n+1}\sin^2[-\pi/2+ \theta_{a_x,b_y}/2^{n}]} \\
& =  \frac{2}{2^{3n+1} \sin^2[\theta_{a_x,b_y}/2^{n}]} + \frac{2}{2^{3n+1}\cos^2[ \theta_{a_x,b_y}/2^{n}]} 
\end{align*}
where $\mathcal{S}_{a_x}=\{a_x, a_x+2^{n-1}\}$, $\mathcal{S}_{b_y}=\{b_y, b_y+2^{n-1}\}$, and $\theta_{a_x, b_y} = \pi (a_x-b_y+ \alpha_x + \beta_y)$. Here, the last equation is reduced by using the double angle formulae to 
\begin{align*}
p_{d=2^{n-1}}^{\text{CGLMP}}(a_x,b_y|x,y) = \frac{1}{2^{3n-2} \sin^2[\theta_{a_x,b_y}/2^{n-1}]},
\end{align*}
which is equal to the right-hand side of Eq. \eqref{eq:merge}. 
}

\section{Numerical data in Fig. \ref{path_3d}}\label{Appdx2}

\begin{figure*}[t!]
  \centering
    \includegraphics[width =7cm]{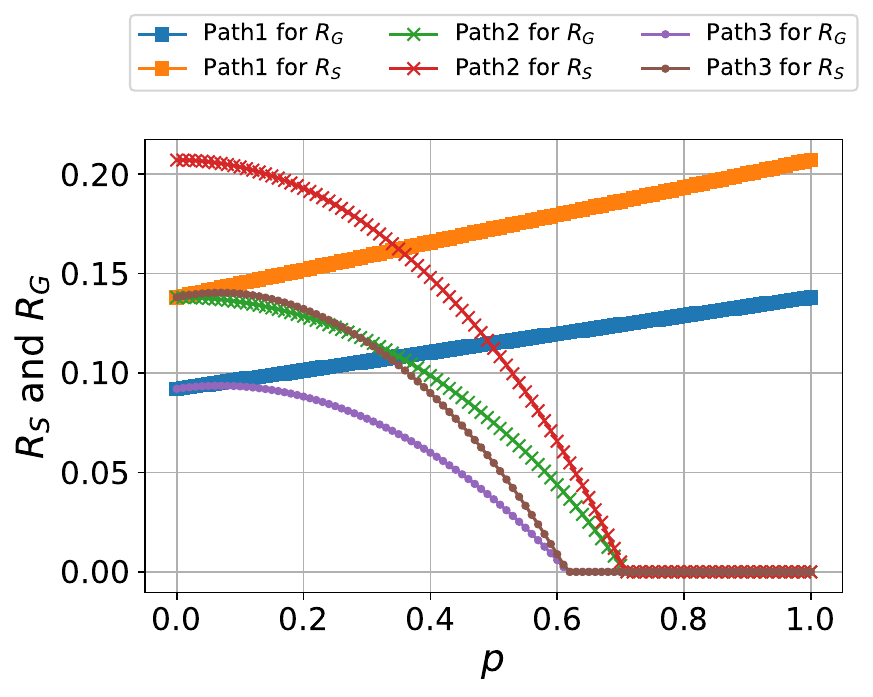}
    \includegraphics[width =7.5cm]{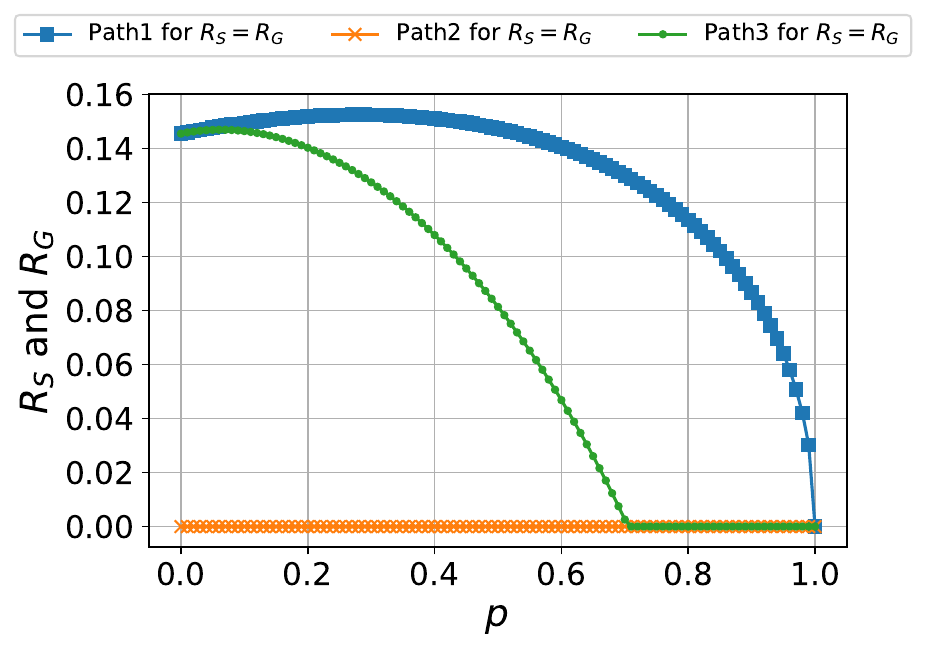}
    
    \;\;\;
    (a) $R_G$ and $R_S$ for $d_\text{eff} = 2$
    \;\;\;\;\;\;\;\;\;\;\;\;\;\;\;\;\;\;\;
    (b) $R_G$ and $R_S$ for $d_\text{eff} = 3$
    \caption{Inequivalence between standard and generalized robustness measures}\label{fig:3_paths}
\end{figure*}

We discover that the inequivalence between the standard and generalized robustness measures arises between the even- and odd-dimensional cases up to $d=d_{\text{eff}}=8$ when measuring the maximally entangled state using the CGLMP settings, as shown in Fig. \ref{Ineq_d_1}. 
This inequivalence is especially prominent for dimensions two and three, warranting further investigation.
We conducted these investigations by systematically varying entangled states along paths 1,2, and 3, as shown in Figs. \ref{R_d3}. The numerical data presented in Fig. \ref{R_d3} were obtained by varying the parameter $p$ within the range $0\geq p\geq 1$ to determine the states for $d_\text{eff} = 2$ as follows:
\begin{align*}
\text{Path 1: }&\sqrt{\frac{1}{3}+\frac{p}{6}}(|00\rangle+|11\rangle) + \sqrt\frac{{1-p}}{ 3}|22\rangle \\
\text{Path 2: }& \sqrt{\frac{1+p}{2}}|00\rangle+\sqrt{\frac{1-p}{2}}|11\rangle\\
\text{Path 3: }&\sqrt{\frac{1+2p}{3}}|00\rangle+ \sqrt{\frac{1-p}{3}}(|11\rangle + |22\rangle)
\end{align*}
Similarly, for $d_\text{eff}=3$, we parametrically varied the states with respect to $p$ as follows:
\begin{align*}
\text{Path 1: }&\sqrt{\frac{1}{3}+\frac{p}{6}}(|00\rangle+|22\rangle) + \sqrt\frac{{1-p}}{ 3}|11\rangle \\
\text{Path 2: }& \sqrt{\frac{1+p}{2}}|00\rangle+\sqrt{\frac{1-p}{2}}|22\rangle\\
\text{Path 3: }&\sqrt{\frac{1+2p}{3}}|00\rangle+ \sqrt{\frac{1-p}{3}}(|11\rangle + |22\rangle).
\end{align*}
These parameterized states encompass the maximal violation point within the trajectories, which is achieved by the partially entangled state $(|00\rangle+\gamma|11\rangle+|22\rangle)/\sqrt{n}$.

Consequently, for $d_\text{eff} = 2$, $R_S$ had 1.5 times larger values than $R_G$. This consistent ratio implies that there is no inequivalence, as shown in Fig. \ref{fig:3_paths} (a). Furthermore, for $d_\text{eff}$, these values were the same. Inequivalent behaviors were not observed when entangled states were varied with fixed measurement settings. However, inequivalence emerged when we compared the robustness measures obtained using different measurement settings, as shown in Fig. \ref{path_3d}.

\bibliographystyle{iopart-num}
\bibliography{mybibfile}

\end{document}